# The Standard Model Without Higgs Bosons in the Fermion Sector

V.P. Neznamov

RFNC-VNIIEF, 607190, Sarov, N.Novgorod region

e-mail: Neznamov@vniief.ru

**Abstract**

The paper addresses the construction of the Standard Model with massive fermions without introduction of the Yukawa interaction between Higgs bosons and fermions. With such approach, Higgs bosons are responsible only for the gauge invariance of the theory's boson sector and interact only with gauge bosons $W^{\pm}, Z$, gluons and photons.

As we know, to provide *SU(2)* – invariance of the theory, the Standard Model first considers massless fermions that are given masses after the mechanism of spontaneous symmetry violation is introduced, Higgs bosons appear and their gauge invariant interaction with Yukawa-type fermions is postulated [1].

On the eve of the decisive *LHC* experiments one can question oneself whether it is possible to construct the Standard Model with initially massive fermions, while preserving the theory's *SU(2)* – symmetry.

In this case, Higgs bosons are responsible only for the gauge invariance of the theory's boson sector and interact only with gauge bosons $W^{\pm}, Z$, gluons and photons.

With the theory defined in this manner, fermion masses are introduced from outside. The theory has no vertices of Yukawa interactions between fermions and Higgs bosons and, therefore, there are no processes of scalar boson decay to fermions $\left(H \rightarrow f \bar{f}\right)$, no quarkonium states $\psi, \Upsilon, \theta$ including Higgs bosons, no interactions of Higgs bosons with gluons ($ggH$) and photons ($\gamma\gamma H$) via fermion loops, etc.

The answer to the question above has already been given in papers [2], [3], where the Standard Model is derived in the modified Foldy-Wouthuysen representation. It has been shown that for its being *SU(2)*-invariant, the theory formulated in the Foldy-Wouthuysen representation does not necessarily require Higgs bosons to interact with fermions, while all theoretical and experimental implications of the Standard Model obtained in the Dirac representation are preserved. The goal of this paper is to construct, in a similar way, the Standard Model with initially massive fermions and spinors in the Dirac representation to meet the requirements of local *SU(3)×SU(2)×U(1)* symmetry.

The paper uses the system of units, where $\hbar = c = 1$; $x, p, B$ are 4-vectors; $p^{\mu} = i\dfrac{\partial}{\partial x_{\mu}}$;

the inner product is taken in the from

$$xy = x^\mu y_\mu = x^0 y^0 - x^k y^k, \mu = 0,1,2,3; k = 1,2,3;$$

$$\alpha^\mu = \begin{cases} 1, \mu = 0 \\ \alpha^k, \mu = k = 1,2,3 \end{cases}; \gamma^\mu = \gamma^0 \alpha^\mu; \beta = \gamma^0, \alpha^k, \gamma^\mu, \gamma^5 - \text{ are Dirac matrices}$$

Consider the density of Hamiltonian of a Dirac particle with mass $m_f$, which interacts with an arbitrary abelian boson field $B^\mu$

$$\mathcal{H}_D = \psi^\dagger (\vec{\alpha}\vec{p} + \beta m_f + q\alpha_\mu B^\mu)\psi = \psi^\dagger (P_L + P_R)(\vec{\alpha}\vec{p} + \beta m_f + q\alpha_\mu B^\mu)(P_L + P_R)\psi =$$

$$= \psi_L^\dagger (\vec{\alpha}\vec{p} + q\alpha_\mu B^\mu)\psi_L + \psi_R^\dagger (\vec{\alpha}\vec{p} + q\alpha_\mu B^\mu)\psi_R + \psi_L^\dagger \beta m_f \psi_R + \psi_R^\dagger \beta m_f \psi_L \quad (1)$$

In (1), $q$ – is the coupling constant; $P_L = \dfrac{1-\gamma_5}{2}, P_R = \dfrac{1+\gamma_5}{2}$ – are the left and right projection operators; $\psi_L = P_L \psi, \psi_R = P_R \cdot \psi$ – are the left and right components of the Dirac field operator $\psi$.

The reason why the abelian case is considered for the field $B^\mu$ is simplicity. As will be shown below, using a general case of a Dirac particle interacting with non-abelian boson fields would not change the conclusions and implications of this paper.

Using the density of Hamiltonian $\mathcal{H}_D$ the motion equations for $\psi_L$ and $\psi_R$ can be obtained:

$$\begin{aligned} p_0 \psi_L &= (\vec{\alpha}\vec{p} + q\alpha_\mu B^\mu)\psi_L + \beta m_f \psi_R \\ p_0 \psi_R &= (\vec{\alpha}\vec{p} + q\alpha_\mu B^\mu)\psi_R + \beta m_f \psi_L \end{aligned} \quad (2)$$

One can see that both the density of Hamiltonian $\mathcal{H}_D$ and motion equations have a form, which is not $SU(2)$ – invariant because of the Dirac fermion having a mass.

It follows from Eqs. (2) that

$$\begin{aligned} \psi_L &= (p_0 - \vec{\alpha}\vec{p} - q\alpha_\mu B^\mu)^{-1} \beta m_f \psi_R \\ \psi_R &= (p_0 - \vec{\alpha}\vec{p} - q\alpha_\mu B^\mu)^{-1} \beta m_f \psi_L \end{aligned} \quad (3)$$

By substituting (3) to the right-hand side of Eqs. (2) proportional to $\beta m_f$, we obtain integro-differential equations for $\psi_R$ and $\psi_L$

$$\left[\left(p_0 - \vec{\alpha}\vec{p} - q(\alpha_0 B^0 - \vec{\alpha}\vec{B})\right) - \beta m_f \left(p_0 - \vec{\alpha}\vec{p} - q(\alpha_0 B^0 - \vec{\alpha}\vec{B})\right)^{-1} \beta m_f\right]\psi_L = 0$$
$$\left[\left(p_0 - \vec{\alpha}\vec{p} - q(\alpha_0 B^0 - \vec{\alpha}\vec{B})\right) - \beta m_f \left(p_0 - \vec{\alpha}\vec{p} - q(\alpha_0 B^0 - \vec{\alpha}\vec{B})\right)^{-1} \beta m_f\right]\psi_R = 0 \quad (4)$$

One can see that equations for $\psi_R$ and $\psi_L$ have the same form, and, in contrast to Eqs. (2), the presence of mass $m_f$ does not lead to mixing the right and left components of $\psi$.

Eqs. (4) can be written as

$$\left[\left(p_0 - \vec{\alpha}\vec{p} - q\alpha_\mu B^\mu\right) - \left(p_0 + \vec{\alpha}\vec{p} - q\bar{\alpha}_\mu B^\mu\right)^{-1} m^2\right]\psi_{L,R} = 0 \quad (5)$$

In expression (5), $\psi_{L,R}$ shows that equations for $\psi_L$ and $\psi_R$ have the same form;

$$\bar{\alpha}_\mu = \begin{cases} 1 \\ -\alpha^i \end{cases}.$$

If we multiply Eqs. (5) on the left side by term $p_0 + \vec{\alpha}\vec{p} - q\bar{\alpha}_\mu B^\mu$, we obtain second-order equations with respect to $p^\mu$

$$\left[\left(p_0 + \vec{\alpha}\vec{p} - q\bar{\alpha}_\mu B^\mu\right)\left(p_0 - \vec{\alpha}\vec{p} - q\alpha_\mu B^\mu\right) - m^2\right]\psi_{L,R} = 0 \quad (6)$$

For the case of quantum electrodynamics $(q = e, B^\mu = A^\mu)$, Eqs. (6) have the form

$$\left[(p_0 - eA_0)^2 - (\vec{p} - e\vec{A})^2 - m^2 + e\vec{\sigma}\vec{H} + i\vec{\alpha}\vec{E}\right]\psi_{L,R} = 0 \quad (7)$$

In Eqs. (7) $\vec{H} = rot\,\vec{A}$ is magnetic field, and $\vec{E} = -\dfrac{\partial \vec{A}}{\partial t} - \nabla A_0$ is electrical field,

$\vec{\sigma} = \begin{pmatrix} \vec{\sigma}' & 0 \\ 0 & \vec{\sigma}' \end{pmatrix}$, $\sigma'^i$ - matrices Pauli.

Eqs. (7) coincide with the second-order equation obtained by Dirac in the 1920s [4]. However, in contrast to [4] (see also [5]), Eqs. (7) contain no "excess" solutions. The operator $\gamma_5$ commutes with Eqs. (6). Consequently, $\gamma_5\psi = \delta\psi\,(\delta^2 = 1; \delta = \pm 1)$. The case of $\delta = -1$ corresponds to the solution of Eq. (7) for $\psi_L$, and $\delta = +1$ corresponds to the solution of Eq. (7) for $\psi_R$.

Eqs. (5), (6) are $SU(2)$-invariant, but they are nonlinear with respect to the operator $p_0 = i\frac{\partial}{\partial t}$. Linear forms of $SU(2)$-invariant equations for fermion fields relative to $p_0$ can be obtained using the Foldy-Wouthuysen transformation [6] in a specially introduced isotopic space.

We now introduce an eight-component field operator, $\Phi_1 = \begin{pmatrix} \psi_R \\ \psi_L \end{pmatrix}$, and isotopic matrices, $\tau_3 = \begin{pmatrix} I & 0 \\ 0 & -I \end{pmatrix}$, $\tau_1 = \begin{pmatrix} 0 & I \\ I & 0 \end{pmatrix}$, acting on the four upper and four lower components of operator $\Phi_1$. So, Eqs. (2) can be written as

$$p_0 \Phi_1 = \left( \vec{\alpha}\vec{p} + \tau_1 \beta m_f + q\alpha_\mu B^\mu \right) \Phi_1 \qquad (8)$$

As $\tau_1$ commutes with the right-hand side of Eq. (8), field $\Phi_2 = \tau_1 \Phi_1 = \begin{pmatrix} \psi_L \\ \psi_R \end{pmatrix}$ is also solution to Eq. (8).

Further, consider Eq. (8) without boson field $B^\mu$ (free motion)

$$p_0 \Phi_{1,2} = \left( \vec{\alpha}\vec{p} + \tau_1 \beta m_f \right) \Phi_{1,2} \qquad (9)$$

$\Phi_{1,2}$ shows that Eqs. (9) are the same for fields $\Phi_1$, $\Phi_2$.

Now we find the Foldy-Wouthuysen transformation in the isotopic space for free motion Eq. (9) using the Eriksen transformation [7].

$$U_{FW}^0 = U_{Er} = \frac{1}{2}(1 + \tau_3 \lambda) \left( \frac{1}{2} + \frac{\tau_3 \lambda + \lambda \tau_3}{4} \right)^{-\frac{1}{2}} \qquad (10)$$

In expression (10), we have $\lambda = \frac{\vec{\alpha}\vec{p} + \tau_1 \beta m_f}{E}$; $E = \left( \vec{p}^2 + m^2 \right)^{\frac{1}{2}}$. Since $\left( \vec{\alpha}\vec{p} + \tau_1 \beta m_f \right)^2 = E^2$, $\lambda^2 = 1$.

Expression (10) can be transformed to obtain the following expression:

$$U_{FW}^0 = U_{Er} = \frac{1}{2} \left( 1 + \frac{\tau_3 \vec{\alpha}\vec{p} + \tau_3 \tau_1 \beta m}{E} \right) \left( \frac{1}{2} + \frac{\tau_3 \vec{\alpha}\vec{p}}{2E} \right)^{-\frac{1}{2}} = \\
= \sqrt{\frac{E + \tau_3 \vec{\alpha}\vec{p}}{2E}} \left( 1 + \frac{1}{E + \tau_3 \vec{\alpha}\vec{p}} \tau_3 \tau_1 \beta m \right) \qquad (11)$$

Expression (11) is a unitary transformation $\left( U_{FW}^0 \left( U_{FW}^0 \right)^\dagger = 1 \right)$, and

$$H_{FW} = U_{FW}^0 \left(\vec{\alpha}\vec{p} + \tau_1 \beta m_f\right)\left(U_{FW}^0\right)^\dagger = \tau_3 E \tag{12}$$

Thus, Eqs. (9) in the Foldy-Wouthuysen representation have the form

$$p_0 \left(\Phi_{1,2}\right)_{FW} = \tau_3 E \left(\Phi_{1,2}\right)_{FW} \tag{13}$$

When converting to the Foldy-Wouthuysen representation, in addition to the condition of the Hamiltonian being block-diagonal (13), one should necessarily meet the requirement that the upper or lower components of the field operators $\Phi_1, \Phi_2$ [8] should be zero. One can term this condition as reduction of fields $\Phi_1, \Phi_2$.

Let us check whether this condition is met in our case, or not. Given Eqs. (2), (3), normalized solutions to Eq. (9) for the field operators $\Phi_1, \Phi_2$ can be expressed as follows:

$$\Phi_1^{(+)}(\vec{x},t) = e^{-iEt} \begin{pmatrix} \psi_R^{(+)}(\vec{x}) \\ \dfrac{1}{E - \vec{\alpha}\vec{p}} \beta m \psi_R^{(+)}(\vec{x}) \end{pmatrix}; \quad \Phi_1^{(-)}(\vec{x},t) = e^{iEt} \begin{pmatrix} -\dfrac{1}{E + \vec{\alpha}\vec{p}} \beta m \psi_L^{(-)}(\vec{x}) \\ \psi_L^{(-)}(\vec{x}) \end{pmatrix}$$

$$\Phi_2^{(+)}(\vec{x},t) = e^{-iEt} \begin{pmatrix} \psi_L^{(+)}(\vec{x}) \\ \dfrac{1}{E - \vec{\alpha}\vec{p}} \beta m \psi_L^{(+)}(\vec{x}) \end{pmatrix}; \quad \Phi_2^{(-)}(\vec{x},t) = e^{iEt} \begin{pmatrix} -\dfrac{1}{E + \vec{\alpha}\vec{p}} \beta m \psi_R^{(-)}(\vec{x}) \\ \psi_R^{(-)}(\vec{x}) \end{pmatrix}$$

(14)

In (14), $\Phi_1^{(+)}, \Phi_2^{(+)}; \Phi_1^{(-)}, \Phi_2^{(-)}$ are solutions with positive and negative energy, respectively.

$$\psi_R^{(+)}(\vec{x}) = \frac{1}{2}(1+\gamma_5)\psi_D^{(+)}(\vec{x}) = \frac{1}{2}\sqrt{\frac{E+m}{2E}}\begin{pmatrix} \left(1+\dfrac{\vec{\sigma}\vec{p}}{E+m}\right)\varphi^{(+)}(\vec{x}) \\ \left(1+\dfrac{\vec{\sigma}\vec{p}}{E+m}\right)\varphi^{(+)}(\vec{x}) \end{pmatrix}$$

$$\psi_R^{(-)}(\vec{x}) = \frac{1}{2}(1+\gamma_5)\psi_D^{(-)}(\vec{x}) = \frac{1}{2}\sqrt{\frac{E+m}{2E}}\begin{pmatrix} \left(1-\dfrac{\vec{\sigma}\vec{p}}{E+m}\right)\chi^{(-)}(\vec{x}) \\ \left(1-\dfrac{\vec{\sigma}\vec{p}}{E+m}\right)\chi^{(-)}(\vec{x}) \end{pmatrix}$$

(15)

$$\psi_L^{(+)}(\vec{x}) = \frac{1}{2}(1-\gamma_5)\psi_D^{(+)}(\vec{x}) = \frac{1}{2}\sqrt{\frac{E+m}{2E}}\begin{pmatrix} \left(1-\dfrac{\vec{\sigma}\vec{p}}{E+m}\right)\varphi^{(+)}(\vec{x}) \\ -\left(1-\dfrac{\vec{\sigma}\vec{p}}{E+m}\right)\varphi^{(+)}(\vec{x}) \end{pmatrix}$$

$$\psi_L^{(-)}(\vec{x}) = \frac{1}{2}(1-\gamma_5)\psi_D^{(-)}(\vec{x}) = \frac{1}{2}\sqrt{\frac{E+m}{2E}}\begin{pmatrix} -\left(1+\dfrac{\vec{\sigma}\vec{p}}{E+m}\right)\chi^{(-)}(\vec{x}) \\ \left(1+\dfrac{\vec{\sigma}\vec{p}}{E+m}\right)\chi^{(-)}(\vec{x}) \end{pmatrix}$$

In expressions (15), $\varphi^{(+)}(\vec{x}), \chi^{(-)}(\vec{x})$ are normalized two-component solutions of the Dirac equation with positive and negative energy. In (14), (15), $E$ and $\vec{p}$ are respective operators.

According to (15),

$$\vec{\alpha}\vec{p}\psi_R^{(+)}(\vec{x}) = \vec{\sigma}\vec{p}\psi_R^{(+)}(\vec{x})\,;\ \vec{\alpha}\vec{p}\psi_R^{(-)}(\vec{x}) = \vec{\sigma}\vec{p}\psi_R^{(-)}(\vec{x})$$

$$\vec{\alpha}\vec{p}\psi_L^{(+)}(\vec{x}) = -\vec{\sigma}\vec{p}\psi_L^{(+)}(\vec{x})\,;\ \vec{\alpha}\vec{p}\psi_L^{(-)}(\vec{x}) = -\vec{\sigma}\vec{p}\psi_L^{(-)}(\vec{x})$$

Expr. (15) lead to the following normalizing conditions:

$$\psi_R^{(+)\dagger}(\vec{x})\psi_R^{(+)}(\vec{x}) = \varphi^{(+)\dagger}(\vec{x})\frac{E+\vec{\sigma}\vec{p}}{2E}\varphi^{(+)}(\vec{x})$$

$$\psi_R^{(-)\dagger}(\vec{x})\psi_R^{(-)}(\vec{x}) = \chi^{(-)\dagger}(\vec{x})\frac{E-\vec{\sigma}\vec{p}}{2E}\chi^{(-)}(\vec{x})$$

$$\psi_L^{(+)\dagger}(\vec{x})\psi_L^{(+)}(\vec{x}) = \varphi^{(+)\dagger}(\vec{x})\frac{E-\vec{\sigma}\vec{p}}{2E}\varphi^{(+)}(\vec{x}) \quad (16)$$

$$\psi_L^{(-)\dagger}(\vec{x})\psi_L^{(-)}(\vec{x}) = \chi^{(-)\dagger}(\vec{x})\frac{E+\vec{\sigma}\vec{p}}{2E}\chi^{(-)}(\vec{x})$$

By applying the transformation matrix $U_{FW}^0$ (11) to $\Phi_1, \Phi_2$ (see (14)) we obtain

$$\Phi_{1FW}^{(+)}(\vec{x},t) = U_{FW}^0 \Phi_1^{(+)}(\vec{x},t) = e^{-iEt}\begin{pmatrix} \sqrt{\dfrac{2E}{E+\vec{\sigma}\vec{p}}}\psi_R^{(+)}(\vec{x}) \\ 0 \end{pmatrix}$$

$$\Phi_{1FW}^{(-)}(\vec{x},t) = U_{FW}^0 \Phi_1^{(-)}(\vec{x},t) = e^{iEt}\begin{pmatrix} 0 \\ \sqrt{\dfrac{2E}{E+\vec{\sigma}\vec{p}}}\psi_L^{(-)}(\vec{x}) \end{pmatrix}$$

(17)

$$\Phi_{2FW}^{(+)}(\vec{x},t) = U_{FW}^0 \Phi_2^{(+)}(\vec{x},t) = e^{-iEt}\begin{pmatrix} \sqrt{\dfrac{2E}{E-\vec{\sigma}\vec{p}}}\psi_L^{(+)}(\vec{x}) \\ 0 \end{pmatrix}$$

$$\Phi_{2FW}^{(-)}(\vec{x},t) = U_{FW}^0 \Phi_2^{(-)}(\vec{x},t) = e^{iEt}\begin{pmatrix} 0 \\ \sqrt{\dfrac{2E}{E-\vec{\sigma}\vec{p}}}\psi_R^{(-)}(\vec{x}) \end{pmatrix}$$

One can see from relations (17) that the reduction condition is fulfilled and the matrix $U_{FW}^0$ is, indeed, the Foldy-Wouthuysen representation for the fields $\Phi_1, \Phi_2$ in the isotopic space we have introduced.

Eqs. (13) allow us to write the density of the free-motion Hamiltonian of fermions with mass $m_f$ as

$$\mathcal{H}_{FW} = (\Phi_1)^{\dagger}_{FW} \tau_3 E (\Phi_1)_{FW} + (\Phi_2)^{\dagger}_{FW} \tau_3 E (\Phi_2)_{FW} = (\Phi_1^{(+)})^{\dagger}_{FW} E (\Phi_1^{(+)})_{FW} - (\Phi_1^{(-)})^{\dagger}_{FW} E (\Phi_1^{(-)})_{FW} +$$

$$+ (\Phi_2^{(+)})^{\dagger}_{FW} E (\Phi_2^{(+)})_{FW} - (\Phi_2^{(-)})^{\dagger}_{FW} E (\Phi_2^{(-)})_{FW} = \left(\psi_R^{(+)}\right)^{\dagger} \frac{2E}{E+\vec{\sigma}\vec{p}} E \psi_R^{(+)} - \left(\psi_L^{(-)}\right)^{\dagger} \frac{2E}{E+\vec{\sigma}\vec{p}} E \psi_L^{(-)} +$$

$$+ \left(\psi_L^{(+)}\right)^{\dagger} \frac{2E}{E-\vec{\sigma}\vec{p}} E \psi_L^{(+)} - \left(\psi_R^{(-)}\right)^{\dagger} \frac{2E}{E-\vec{\sigma}\vec{p}} E \psi_R^{(-)} \qquad (18)$$

One can see that Hamiltonian (18) is *SU(2)*-invariant, regardless of whether the fermions are massive or massless. Expression (18) shows that two fermion field operators, $(\Phi_1)_{FW}, (\Phi_2)_{FW}$, need to be used to provide a complete description of the free motion of the right and left fermions. Given (15), the density of Hamiltonian (18) bracketed between two-component spinors $\varphi^{(+)}(\vec{x})$ and $\chi^{(-)}(\vec{x})$ has a form that is commonly used in the field theory,

$$\mathcal{H}_{FW} = 2\left(\varphi^{(+)\dagger} E \varphi^{(+)} - \chi^{(-)\dagger} E \chi^{(-)}\right).$$

In the presence of boson fields $B^\mu(x)$ interacting with fermion fields $\Phi_1(x), \Phi_2(x),$ the Foldy-Wouthuysen transformation and Hamiltonian of Eq. (8) in the Foldy-Wouthuysen representation in the isotopic space can be obtained as a series in powers of the coupling constant using the algorithm described in Refs. [2], [9].

As a result, using denotations from Refs. [2], [9], we obtain

$$U_{FW} = U^0_{FW} \left(1 + \delta_1 + \delta_2 + \delta_3 + ...\right) \qquad (19)$$

$$p_0 (\Phi_{1,2})_{FW} = H_{FW} (\Phi_{1,2})_{FW} = \left(\tau_3 E + qK_1 + q^2 K_2 + q^3 K_3 + ...\right)(\Phi_{1,2})_{FW} \qquad (20)$$

The expressions for operators $C$ and $N$ constituting the basis for the interaction Hamiltonian in the Foldy-Wouthuysen representation obtained using the technique of Refs. [2], [9] can be written in the following form in our case:

$$C = \left[U^0_{FW} q\alpha_\mu B^\mu \left(U^0_{FW}\right)^{\dagger}\right]^{even} = qR\left(B^0 - LB^0 L\right)R - qR\left(\vec{\alpha}\vec{B} - L\vec{\alpha}\vec{B}L\right)R$$

$$N = \left[U^0_{FW} q\alpha_\mu B^\mu \left(U^0_{FW}\right)^{\dagger}\right]^{odd} = qR\left(LB^0 - B^0 L\right)R - qR\left(L\vec{\alpha}\vec{B} - \vec{\alpha}\vec{B}L\right)R \qquad (21)$$

$$R = \sqrt{\frac{E+\tau_3\vec{\alpha}\vec{p}}{2E}}; \quad L = \frac{1}{E+\tau_3\vec{\alpha}\vec{p}} \tau_3\tau_1\beta m$$

The superscripts *even, odd* in (21) show the even and odd parts of the operators relative to the upper and lower isotopic components of $\Phi_1$ and $\Phi_2$.

For Eqs. (20), the Hamiltonian density for fermion fields $(\Phi_1)_{FW}, (\Phi_2)_{FW}$, interacting with boson field $B^\mu(x)$ can be written as

$$\mathcal{H}_{FW} = (\Phi_1)^\dagger_{FW} \left( \tau_3 E + qK_1 + q^2 K_2 + q^3 K_3 + ... \right)(\Phi_1)_{FW} +$$

$$+ (\Phi_2)^\dagger_{FW} \left( \tau_3 E + qK_1 + q^2 K_2 + q^3 K_3 + ... \right)(\Phi_2)_{FW} \qquad (22)$$

The expression for the Foldy-Wouthuysen Hamiltonian in parentheses in equation (22) is, by definition, diagonal with respect to the upper and lower components $(\Phi_1)_{FW}, (\Phi_2)_{FW}$ [6], [8], [9].

When solving applied problems in the quantum field theory using the perturbation theory, fermion fields are expanded in solutions of Dirac equations for free motion or for motion in static external fields. In our case, in the Foldy-Wouthuysen representation, we can also expand fermion fields over the basis (17) or over a similar basis of solutions of the Foldy-Wouthuysen equations in static external fields. Then, Hamiltonian density (22) can be expressed through the functions (17), and it is obvious that this expression, similarly to (18), will be $SU(2)$-invariant due to the diagonality.

Thus, expression (22) and Eqs. (20) are invariant relative to $SU(2)$-transformations regardless of fermions having or not having masses.

Formula (22) demonstrates the necessity of using two fermion fields, $(\Phi_1)_{FW}(x), (\Phi_2)_{FW}(x)$, in the formalism for constructing the Standard Model. If only $(\Phi_1)_{FW}(x)$ is used in the theory, motion and interactions of the right fermions, as well as motion and interactions of the left anti-fermions remain. If, on the contrary, only $(\Phi_2)_{FW}(x)$ is used in the theory, motion and interaction of the left fermions, as well as motion and interactions of the right antifermions remain.

More careful analysis shows that even with two fields, $\Phi_{1FW}(x), \Phi_{2FW}(x)$, Hamiltonian (22) contains no interactions between real particles and anti-particles. This happens due to the spinor structure of expressions (17) in the introduced isotopic space. The theory's special feature in the Foldy-Wouthuysen representation is that the Hamiltonian terms include interaction $K_n$ (except $K_1$) of an even number of odd operators $N$ that couple states with positive and negative energy. Therefore, interactions between particles and antiparticles can occur only between real and intermediate virtual states [2], [9]. In order to introduce interactions between real particles and antiparticles into the theory in Refs. [2], [9], the Foldy-Wouthuysen representation had to be modified.

To solve the same problems in our case, let us use the following approach.

Write Eq. (8) for field $\Phi_1(x)$ and the same equation for field $\Phi_2(x)$ in the equivalent form

$$p_0 \Phi_1(x) = (\vec{\alpha}\vec{p} + \tau_1 \beta m_f)\Phi_1(x) + \frac{1}{2}q\alpha_\mu B^\mu \Phi_1(x) + \frac{1}{2}q\alpha_\mu B^\mu \tau_1 \Phi_2(x)$$

(23)

$$p_0 \Phi_2(x) = (\vec{\alpha}\vec{p} + \tau_1 \beta m_f)\Phi_2(x) + \frac{1}{2}q\alpha_\mu B^\mu \Phi_2(x) + \frac{1}{2}q\alpha_\mu B^\mu \tau_1 \Phi_1(x)$$

Eq. (23) uses the equality $\Phi_2(x) = \tau_1 \Phi_1(x)$.

Further, performing the Foldy-Wouthuysen transformation (19) for Eqs. (23), we obtain equations for fields $\Phi_{1FW}(x), \Phi_{2FW}(x)$.

$$p_0 \Phi_{1FW}(x) = \tau_3 E \Phi_{1FW}(x) + \left(\frac{q}{2}K_1 + \left(\frac{q}{2}\right)^2 K_2 + \left(\frac{q}{2}\right)^3 K_3 + ....\right)\Phi_{1FW}(x) +$$

$$+ \left(\frac{q}{2}K_{1\tau_1} + \left(\frac{q}{2}\right)^2 K_{2\tau_1} + \left(\frac{q}{2}\right)^3 K_{3\tau_1} + ...\right)\Phi_{2FW}(x)$$

(24)

$$p_0 \Phi_{2FW}(x) = \tau_3 E \Phi_{2FW}(x) + \left(\frac{q}{2}K_1 + \left(\frac{q}{2}\right)^2 K_2 + \left(\frac{q}{2}\right)^3 K_3 + ...\right)\Phi_{2FW}(x) +$$

$$+ \left(\frac{q}{2}K_{1\tau_1} + \left(\frac{q}{2}\right)^2 K_{2\tau_1} + \left(\frac{q}{2}\right)^3 K_{3\tau_1} + ...\right)\Phi_{1FW}(x)$$

Denotations $K_{n\tau_1}$ in formulas (24) mean that the matrix $\tau_1$ is placed in the operators $C$, $N$ (21) next to the fields $B^\mu$:

$$C_{\tau_1} = qR\left(\tau_1 B^0 - L\tau_1 B^0 L\right) - qR\left(\tau_1\vec{\alpha}\vec{B} - L\tau_1\vec{\alpha}\vec{B}L\right)R$$

$$N_{\tau_1} = qR\left(L\tau_1 B^0 - \tau_1 B^0 L\right) - qR\left(L\tau_1\vec{\alpha}\vec{B} - \tau_1\vec{\alpha}\vec{B}L\right)R \qquad (25)$$

Equations (24) correspond to the Hamiltonian density

$$\mathscr{H}_{FW} = \Phi^\dagger_{1FW}\left(\tau_3 E + \frac{q}{2}K_1 + \left(\frac{q}{2}\right)^2 K_2 + \left(\frac{q}{2}\right)^3 K_3 + ....\right)\Phi_{1FW} +$$

$$+ \Phi^\dagger_{1FW}\left(\frac{q}{2}K_{1\tau_1} + \left(\frac{q}{2}\right)^2 K_{2\tau_1} + \left(\frac{q}{2}\right)^3 K_{3\tau_1} + ....\right)\Phi_{2FW} + \qquad (26)$$

$$+ \Phi^\dagger_{2FW}\left(\tau_3 E + \frac{q}{2}K_1 + \left(\frac{q}{2}\right)^2 K_2 + \left(\frac{q}{2}\right)^3 K_3 + ....\right)\Phi_{2FW} +$$

$$+ \Phi^\dagger_{2FW}\left(\frac{q}{2}K_{1\tau_1} + \left(\frac{q}{2}\right)^2 K_{2\tau_1} + \left(\frac{q}{2}\right)^3 K_{3\tau_1} + ...\right)\Phi_{1FW}$$

By analogy with [2], [9], Feynman rules for calculating specific physical processes in the quantum theory of interacting fields using perturbation theory methods can be derived using Hamiltonian density (26) and Eqs. (24).

The isotopic space we have introduced allows constructing the $SU(2)$-invariant Standard Model with massive fermions. In case of interaction with gauge fields $B_\mu$, the Lagrangian with covariant derivative $D^\mu = \partial^\mu - iqB^\mu$ and fermion fields $\Phi_1 = \begin{pmatrix}\psi_R\\\psi_L\end{pmatrix}, \Phi_2 = \begin{pmatrix}\psi_L\\\psi_R\end{pmatrix}$ can be written as

$$\mathscr{L} = \bar{\Phi}_1 \gamma_\mu D^\mu \Phi_1 - \bar{\Phi}_1 \tau_1 m_f \Phi_1 + \bar{\Phi}_2 \gamma_\mu D^\mu \Phi_2 - \bar{\Phi}_2 \tau_1 m_f \Phi_2 \qquad (27)$$

Using this Lagrangian, the motion equations for fermion fields $\Phi_1, \Phi_2$ with mass $m_f$ (see (23)) can be obtained.

Using the isotopic Foldy-Wouthuysen transformation (11), (19) one can obtain the $SU(2)$-invariant Hamiltonian density (26) and motion equations for fermion fields (24) with appropriate definitions of operators $C_{\tau_1}, N_{\tau_1}, K_{1\tau_1}, K_{2\tau_1}, K_{3\tau_1}...$ (see (25)).

To construct the Standard Model using Hamiltonian density (26), one must replace, as it is done in [2], [3], the interaction vertex $q\alpha_\mu B^\mu$ in the operators $K_1, K_2 ..., K_{1\tau_1}, K_{2\tau_1} ...$ with the interaction vertices of the Standard Model [1].

$$q\alpha_\mu B^\mu \rightarrow eQ_f \alpha^\mu A_\mu + \frac{g_2}{\cos\theta_W}\left[\left(T_f^3 - Q_f \sin^2\theta_W\right)\alpha^\mu\left(\frac{I-\gamma_5}{2}\right) - Q_f \sin^2\theta_W \alpha^\mu\left(\frac{I+\gamma_5}{2}\right)\right]Z_\mu +$$

$$+ \frac{g_2}{\sqrt{2}}\left\{\left[(f=u)\alpha_\mu\left(\frac{I-\gamma_5}{2}\right)(f=d) + (f=v_e)\alpha^\mu\left(\frac{I-\gamma_5}{2}\right)(f=e)\right]W_\mu^+ + Hermit.conj.\right\} +$$

$$+ \frac{g_3}{2}\left[(f=u,d)_\alpha \alpha^\mu \lambda^a_{\alpha\beta}(f=u,d)_\beta G^a_\mu\right] \qquad (28)$$

In (28), $A_\mu$ is the electromagnetic field; $Z_\mu, W_\mu^\pm$ are the gauge boson fields; $G^a_\mu$ are the gluon fields; $Q_f$ is the fermion electric charge in the units of $e$; $T_f^3 = 1/2$ for $f = v_e, u$; $T_f^3 = -1/2$ for $f = e, d$; $\theta_W$ is the electroweak mixing angle; $g_2 = \frac{e}{\sin\theta_W}$; $g_3$ is the quantum chromodynamics coupling constant; $\lambda^a$ are generators of group $SU(3)$.

Expressions (28) is written out only for the first lepton and quark family. For the second and the third families it is required to make appropriate substitutions $(v_e, e, u, d) \rightarrow (v_\mu, \mu, c, s)$ and $(v_\tau, \tau, t, b)$ and introduce the quark mixing.

In (28) notations (*f=u*), (*f=d*), etc., imply that spinor *FW*-fields of associated fermions will be located at specified places in the Hamiltonian.

The resulting Standard Model in the Fodly-Wouthuysen representation preserves the *SU(3)* × *SU(2)* ×*U(1)*-invariance and all its theoretical and experimental implications with no interactions required between Higgs bosons and fermions. In this case, Higgs bosons are responsible only for the gauge invariance of the theory's boson sector and interact only with gauge bosons $W^\pm, Z$, gluons and photons.

The suggested version of the Standard Model is, most likely, renormalizable, because the theory's boson sector remains massless till the Higgs spontaneous symmetry violation mechanism is introduced, as quantum electrodynamics with a massless photon and massive electron and positron is a renormalizable theory.

Nevertheless, the question of whether or not the suggested version of the Standard Model is renormalizable needs to be studied more profoundly.

Of course, the results of forthcoming experiments on searching for scalar bosons using the CERN's Large Hadron Collider would provide direct verification of the conclusions of this paper concerning the construction of the Standard Model without interactions between Higgs bosons and fermions.